\documentstyle[a4,12pt]{article}
\begin{document}
\makeatletter
\def\fmslash{\@ifnextchar[{\fmsl@sh}{\fmsl@sh[0mu]}}
\def\fmsl@sh[#1]#2{%
  \mathchoice
    {\@fmsl@sh\displaystyle{#1}{#2}}%
    {\@fmsl@sh\textstyle{#1}{#2}}%
    {\@fmsl@sh\scriptstyle{#1}{#2}}%
    {\@fmsl@sh\scriptscriptstyle{#1}{#2}}}
\def\@fmsl@sh#1#2#3{\m@th\ooalign{$\hfil#1\mkern#2/\hfil$\crcr$#1#3$}}
\makeatother
\global\arraycolsep=2pt 
\thispagestyle{empty}
\begin{titlepage}

\begin{flushright}
hep-ph/9411370\\
HUB--IEP--94/25
\end{flushright}

$\vphantom{a}$

\vspace{1cm}
\vfill
\begin{center}
\Large\bf Combining Vector-- and  Heavy Quark Symmetry
\end{center}

\vspace{0.8cm}

\begin{center}
\renewcommand{\thefootnote}{\fnsymbol{footnote}}
Thorsten Feldmann\footnote{supported by
{\it Deutsche Forschungsgemeinschaft}}  \\
{\sl Institut f\"ur Physik, Humboldt--Universit\"at zu Berlin, \\
     Invalidenstra\ss e 110, D--10115 Berlin, Germany }
\renewcommand{\thefootnote}{\arabic{footnote}}
\setcounter{footnote}{0}
\vspace*{5mm}  \\
Thomas Mannel  \\
{\sl Institut f\"{u}r Theoretische Teilchenphysik,
     Universit\"{a}t Karlsruhe, \\
     Kaiserstra{\ss}e 12,
     D -- 76128 Karlsruhe, Germany.}
\end{center}

\vspace{5.0cm}

\begin{abstract}
\noindent
Assuming that the chiral symmetry of the light degrees of freedom
is realized as the vector symmetry proposed by Georgi, we write
down a chiral Lagrangian for heavy mesons which incorporates both
heavy quark and vector symmetry.  Some of the phenomenological
implications of this idea are considered.
\end{abstract}
\vfill
\centerline{(submitted to Physics Letters B)}
\vfill
\end{titlepage}

\setcounter{page}{1}
\section{Introduction}
The combination of heavy quark and chiral symmetries has turned out
to be a powerful tool to obtain model independent predictions for
various processes. In particular, the decays of a heavy meson into
light pseudoscalars has been considered in this framework and some
interesting predictions for strong processes as well as for
electromagnetic transitions have been obtained \cite{chiral}.

This approach to heavy to light decays has been extended to light
vector mesons by using the approach of hidden symmetry \cite{chivec}.
In this approach the light vector mesons are introduced as gauge bosons
of a hidden symmetry which is broken spontaneously to give masses
to the light vector mesons \cite{hidden,hidden1,hidden2}.
{In this way one may implement chiral and Lorentz invariance in an
elegant way; it has been shown \cite{hidden} that writing down all couplings
allowed by these two symmetries \cite{We68} yields the hidden symmetry
Lagrangian. Including heavy mesons along the lines proposed in
\cite{chivec}, the resulting heavy-meson chiral Lagrangian has, however,
a large number of unknown coupling parameters, even to lowest order in
the chiral expansion.}

It has has been pointed out by Georgi \cite{VL} that the hidden symmetry
method has an interesting limit in which additional symmetries
occur. In this limit, the so called vector limit, the chiral
symmetry is in fact realized in an unbroken way,
the vector meson octett becomes massless and the scalar fields
corresponding to the longitudinal components of the vecor mesons
become the chiral partners of the pions. The phenomenology of this
model has been worked out in some detail \cite{Cho}.

In fact, there are arguments that this realization of chiral symmetry
may be obtained from the dynamics of QCD. Of course, it is impossible
to trace what happens to chiral symmetry once the scales become that
low that one enters the nonperurbative region of QCD, but at least
the enlarged symmetry of the
vector limit, the so called vector symmetry, is compatible
with the $N_c \to \infty$ limit of QCD, where $N_c$ is the number of
colors \cite{VL}. Another way such a limit may be realized is
the scenario of ``mended symmetries'' \cite{We90}, which is
envisaged to hold  sufficiently
close to the chiral symmetry restoration point. Further arguments
in favour the vector limit have been given recently \cite{VLQCD}.

We shall adopt the point of view that the dynamics of QCD is indeed
such that reality is sufficiently close to the vector limit and that
it may be used as a starting point. In that case the enlarged symmetry
in the vector limit yields a reduction of the number of independent
coupling constants, which may appear in the chiral Lagrangian at low
energies. In fact, in the vector limit the dynamics of the
light degrees of freedom (in this
case the pions and the longitudinal components of the light vector mesons)
is determined solely by $f_\pi$, the pion decay constant.

However, in order to have a good description of data one has to
take into account the breaking of vector symmetry.
For instance, the relation obtained for the mass of the $\rho$
meson is a factor $\sqrt{2}$ off from the usual KSRF relation,
yielding a $\rho$ mass which is a factor of $\sqrt{2}$ too large.
This factor indicates the size of corrections one has to expect
when using the vector limit. Other examples are the application of
the vector limit to hadronic decays of $D$ mesons \cite{VLHQ} and
to intrinsic parity violating decays \cite{VLWZ},
where an agreement of similar quality has been observed.

The coupling of the transverse components of the $\rho$ meson
breaks the enlarged symmetry, and the above prediction for the
$\rho$ mass is obtained, if this coupling is the only source of
symmetry breaking. The situation
may be improved by including systematically other sources
of vector symmetry breaking. This has been studied in detail for the
hadronic decays of the $\tau$ lepton \cite{VLtau} and for the
hadronic $D$ meson decays \cite{VLHQ1,VLHQ2}.

The present note extends the combination of heavy quark and
chiral symmetry to the case where the symmetry of the light
degrees of freedom is the vector symmetry. Due to the enlarged
symmetry relations between the coupling constants are obtained, which
reduce the number of independent coupling constant to only one
in the vector limit. However, corrections to the vector limit
may be sizable and the predictions of the vector limit may
obtain corrections factors similar to the one for
the $\rho$ mass, namely a factor of $\sqrt{2}$. In the present
note, we shall not consider any symmetry breaking terms
but shall explore only the consequences of the symmetry limit.

The paper is organized as follows. In the next section we shall
recall the concept of hidden and vector symmetry of the light
degrees of freedom. In section 3 the coupling of the heavy
mesons to the light degrees of freedom is considered and the
relations between coupling constants in the vector limit are
given. Finally, we shall discuss some phenomenological implications
and conclude.

\section{Hidden Symmetry and the Vector Limit}
The chirally symmetric Lagrangian for the light pseudoscalars
is conveniently obtained from a
nonlinear representation of the chiral $SU(3)_L
\otimes SU(3)_R$ symmetry \cite{CCWZ}. It is defined  in terms
of a matrix field $\xi$
\begin{equation}
\xi = \exp \left( \frac{i}{f} \pi \right) \quad
\pi = \pi^a T^a \quad \mbox{Tr }(T^a T^b) = \frac{1}{2} \delta^{ab}
\end{equation}
where $T^a$ denote the generators of $SU(3)$ in the fundamental
representation. Under chiral $SU(3)_L \otimes SU(3)_R$ the field
$\xi$ transforms as
\begin{equation} \label{nlrep}
\xi \to L \xi U^\dagger = U \xi R^\dagger \quad
L \in SU(3)_L \quad
R \in SU(3)_R
\end{equation}
The Lagrangian to lowest order (i.e. containing the lowest number
of derivatives of the fields) for the light pseudoscalar octett is
as usual
\begin{equation}
{\cal L}_0 = \frac{f^2}{4} \mbox{Tr } \left\{
(\partial_\mu \Sigma)^\dagger
(\partial^\mu \Sigma) \right\}
\mbox{ with } \Sigma = \xi^2 ,
\end{equation}
where $f = 93$ MeV is the pion decay constant.

To include also the light vector mesons the so called hidden symmetry
approach has been proposed in \cite{hidden}. This method amounts
to insert an additional ``hidden'' $SU(3)_H$ by defining matrix
fields
$\Sigma_L$ and $\Sigma_R$ which transform as
\begin{eqnarray}
&& \Sigma_L \to L \Sigma_L h^\dagger \quad L \in SU(3)_L
    \quad h \in SU(3)_H \\
&& \Sigma_R \to R \Sigma_R h^\dagger \quad R \in SU(3)_R
    \quad h \in SU(3)_H
\end{eqnarray}
where $SU(3)_H$ is the so called hidden symmetry. The light vector
mesons are introduced as gauge fields of the local hidden $SU(3)_H$;
since this group is spontaneously broken, the vector mesons will
aquire a mass. This becomes evident by expressing the two fields
$\Sigma_{L/R}$ in terms of the field $\xi$
\begin{eqnarray} \label{SL}
\Sigma_L &=& \xi \exp \left(\frac{i}{f} s \right) \quad s = s^a T^a
\\ \label{SR}
\Sigma_R &=& \xi^\dagger \exp \left(\frac{i}{f} s \right)
\end{eqnarray}
The additional scalar fields $s$ are the Goldstone bosons of the
broken hidden symmetry and will become the longitudinal components
of the vector mesons.

The Lagrangian for the light pseudoscalars and the light vector mesons
is conveniently formulated in terms of the currents
\begin{eqnarray}
{\cal A}_\mu &=& \frac{i}{2} \left\{
\Sigma_R^\dagger \partial_\mu \Sigma_R -
\Sigma_L^\dagger \partial_\mu \Sigma_L \right\} \quad
{\cal A}_\mu \to h {\cal A}_\mu h^\dagger
\\
{\cal V}_\mu &=& \frac{i}{2} \left\{
\Sigma_R^\dagger\partial_\mu\Sigma_R   +
\Sigma_L^\dagger\partial_\mu\Sigma_L   \right\} \quad
{\cal V}_\mu \to h {\cal V}_\mu h^\dagger + i h \partial_\mu h^\dagger
\end{eqnarray}
and reads
\begin{eqnarray} \label{L01}
{\cal L}_{01} &=& -f^2 \mbox{Tr } \left\{
{\cal A}_\mu {\cal A}^\mu \right\}
- \frac{1}{2} \mbox{Tr } \left\{ F_{\mu \nu} F^{\mu \nu} \right\}
\\
&-& a f^2 \mbox{Tr } \left\{
\left[ {\cal V}_\mu - g_V \rho_\mu \right]
\left[ {\cal V}^\mu - g_V \rho^\mu \right] \right\}
\nonumber
\end{eqnarray}
where $\rho = \rho^a T^a$ are the fields of the light vector mesons.
Under the hidden symmetry it transforms like a gauge field
\begin{equation}
\rho_\mu \to h \rho_\mu h^\dagger
+ \frac{i}{g_V} h \partial_\mu h^\dagger .
\end{equation}
$F_{\mu \nu}$ is the ususal field strength tensor
\begin{equation}
F_{\mu \nu} = \partial_\mu \rho_\nu
            - \partial_\nu \rho_\mu + g_V [\rho_\mu , \rho_\nu ] .
\end{equation}
This Lagrangian contains (aside from the pion decay constant $f$)
the coupling constant $g_V$ and the parameter $a$. These two
parameters are fixed by the value of the coupling strength of the
$\rho$ to the pions $g_{\rho \pi \pi}$ and the mass of the $\rho$
meson. One obtains for the two parameters in terms of the input
\begin{equation}
g_{\rho \pi \pi} = \frac{a}{2} g_V \qquad
m_\rho^2 = \frac{4}{a} f^2 g_{\rho \pi \pi}^2
\end{equation}
The choice of $a = 2$ yields the KSRF relation
$m_\rho^2 = 2 f^2 g_{\rho \pi \pi}$ \cite{KSRF}
which works surprisingly well, while the gauge coupling is fixed
for $a=2$ to be $g_V = g_{\rho \pi \pi} \sim 6.0$.

It has been pointed out by Georgi \cite{VL} that for the choice
$a=1$ and in the limit $g_V \to 0$,
the symmetry of the system
becomes enlarged to the so called vector symmetry.
The Lagrangian becomes in this limit
\begin{eqnarray}
{\cal L}_{VL} &=& - f^2 \mbox{Tr } \left\{
{\cal A}_\mu {\cal A}^\mu  +
{\cal V}_\mu {\cal V}^\mu \right\} \\
&=& \frac{f^2}{2} \mbox{Tr } \left\{
(\partial_\mu \Sigma_L)(\partial^\mu \Sigma_L)^\dagger +
(\partial_\mu \Sigma_R)(\partial^\mu \Sigma_R)^\dagger \right\}
\end{eqnarray}
which has a global
$$
\frac{SU(3)_L \otimes SU(3)_{H_L} \otimes SU(3)_R \otimes SU(3)_{H_R}}
     {SU(3)_{L+H_L} \otimes SU(3)_{R+H_R}}
$$
symmetry which acts in the following way on the fields
\begin{eqnarray}
&& \Sigma_L \to L \Sigma_L h_L^\dagger \quad L \in SU(3)_L
   \quad h_L \in SU(3)_{H_L} \\
&& \Sigma_R \to R \Sigma_R h_R^\dagger \quad R \in SU(3)_R
   \quad h_R \in SU(3)_{H_R} .
\end{eqnarray}
This means in particular, that for the light degrees of freedom there
is an unbroken chiral symmetry with parity doublets consisting of
the pions and the longitudinal components of the light vector mesons.
This enlarged symmetry is broken by the coupling of the transverse
components of the vector mesons
which act as gauge bosons of the diagonal group $H_L + H_R$. The
resulting Lagrangian is the same as (\ref{L01}) with the choice
$a = 1$.
However, if this coupling is the only source of vector symmetry breaking
one obtains a mass for the vector mesons,
which is a factor 1.4 too large. This indicates that there have to
be additional sources of vector symmetry breaking, which will
contribute substantially; still the limiting case may be of some use
to obtain an idea of the size of the coupling constants considered
below.

\section{Coupling of Heavy Mesons}
We are now ready to consider the coupling of the heavy mesons to
the light pseudoscalar and vector mesons. We shall start by writing
down the Lagrangian for light pseudoscalar and vector mesons for the
general case, including the coupling of these light states to the
heavy ground state spin symmetry doublet, and to an excited spin
symmetry doublet, consisting of a $0^+$ and a $1^+$ state.
We combine the two heavy spin symmetry doublets into
two fields \cite{chiral}
\begin{eqnarray}
H(v) &=& \frac{\sqrt{m_H}}{2} (1 + \fmslash{v})
\left( - \gamma_5 \overline{P}
+ \overline{P}_\mu^* \gamma^\mu \right) \quad \overline{P}^* \cdot v =
0
\\ && \mbox{for the } (0^-, 1^-) \mbox{ spin symmetry doublet}
\nonumber \\
K(v) &=& \frac{\sqrt{m_H}}{2} (1 + \fmslash{v})
\left( - \overline{S} + \overline{S}_\mu^*
       \gamma^\mu \gamma_5 \right) \quad \overline{S}^* \cdot v = 0
\\ && \mbox{for the } (0^+, 1^+) \mbox{ spin symmetry doublet}
\nonumber
\end{eqnarray}
which transfrom as
\begin{equation}
   H(v) \to H(v) h^\dagger \qquad K(v) \to K(v) h^\dagger
\end{equation}
under $SU(3)_L \otimes SU(3)_R \otimes SU(3)_H$.

In leading order the most general Lagrangian consistent with
hidden symmetry and spin symmetry is then given by
\begin{eqnarray} \label{Gatto}
{\cal L}_2 &=&
{\rm Tr} \left[
	\bar{H}^\alpha (i v \cdot D^{\beta\alpha}) H^\beta
	\right]
-{\rm Tr} \left[
	\bar{K}^\alpha (i v \cdot D^{\beta\alpha}) K^\beta
	\right]
\nonumber \\ &&
+ g_H {\rm Tr} \left[
	\bar{H}^\alpha H^\beta \gamma_5
	\fmslash{\cal{A}}^{\beta \alpha} \right]
\; + g_K {\rm Tr} \left[
	\bar{K}^\alpha K^\beta \gamma_5
	\fmslash{\cal{A}}^{\beta \alpha} \right]
\nonumber \\ &&
+ g_H' {\rm Tr} \left[
	\bar{H}^\alpha H^\beta
	(\fmslash{{\cal V}}^{\beta \alpha} -
	 g_V \fmslash{\rho}^{\beta \alpha}) \right]
\; + g_K' {\rm Tr} \left[
	\bar{K}^\alpha K^\beta
	(\fmslash{{\cal V}}^{\beta \alpha}  -
	 g_V \fmslash{\rho}^{\beta \alpha}) \right]
\nonumber \\ &&
+ \lambda_V {\rm Tr} \left[
	\bar{H}^\alpha K^\beta
	(\fmslash{{\cal V}}^{\beta \alpha}
	 - g_V \fmslash{\rho}^{\beta \alpha}) \right]
\; + h.c.
\nonumber \\ &&
+ \lambda_A {\rm Tr} \left[
	\bar{H}^\alpha K^\beta
	\fmslash{\cal{A}}^{\beta \alpha} \right] \; + h.c.
\label{l2}
\end{eqnarray}
where we have written the $SU(3)_H$ indices explicitly,

$D_\mu^{\alpha\beta} = \delta^{\alpha\beta} \partial_\mu
+ i g_V \rho^{\alpha\beta}$ is a covariant derivative of $SU(3)_H$
and the trace is to be taken with respect to the dirac matrix structure.
This is the Lagrangian as it was
written down in \cite{chivec}. It is expressed in terms of
six independent couplings
$g_H,g_K,g_H',g_K',\lambda_V,\lambda_A$, which need to be fixed from
some experimental input.

In the vector limit the chiral
$SU(3)_L \otimes SU(3)_R$ symmetry is unbroken and hence chiral
partners of the heavy particles have to show up. Obviously one
may not use a similar construction as for the light states,
since it makes no sense to consider the massless limit of the heavy
states. The way we shall proceed here is to introduce the chiral partners
directly, i.e. we shall combine the $0^-$ and the $0^+$ state into a
parity doublet, while the $1^-$ and the $1^+$ state form another parity
doublet. From the heavy quark limit it follows that the
splitting between spin symmetry partners is of order $\Lambda_{QCD}$,
while from heavy quark symmetry alone one expects that the splitting
between the $0^-$ and the $0^+$ state should be a constant, which is
independent of the heavy quark mass and related to vector symmetry
breaking. Thus in the combined heavy
quark and vector limit, all four states should be degenerate.

It is convenient to define left and right handed heavy fields as
\begin{eqnarray}
H_L &=& (H + K) \frac{1-\gamma_5}{2} \\
H_R &=& (H + K) \frac{1+\gamma_5}{2}
\end{eqnarray}
which transform under
$SU(3)_R \otimes SU(3)_{H_R} \otimes SU(3)_L \otimes SU(3)_{H_L}$
as
\begin{equation}
   H_L \to H_L h_L^\dagger \qquad H_R \to H_R h_R^\dagger
\end{equation}

The most general lagragian which embodies the enlarged
symmetry of the vector limit is then simply given by
\begin{eqnarray}
{\cal L}_2^{VL} &=&
- {\rm Tr} \left[
	\bar{H}_L^\alpha (i v \cdot \partial) H_L^\alpha \fmslash{v}
	\right]
\;
- {\rm Tr} \left[
	\bar{H}_R^\alpha (i v \cdot \partial) H_R^\alpha \fmslash{v}
	\right]
\nonumber \\ &&
+ g {\rm Tr}
\left[ \bar{H}^\alpha_L H^\beta_L
	(\fmslash{\cal{V}}^{\beta\alpha}
	  - \fmslash{\cal{A}}^{\beta\alpha}) \right]
\;
+ g {\rm Tr}
\left[ \bar{H}^\alpha_R H^\beta_R
	(\fmslash{\cal{V}}^{\beta\alpha}
	  + \fmslash{\cal{A}}^{\beta\alpha}) \right]
\label{l2VL}
\end{eqnarray}
where $g$ is given by the $H^* H \pi$ coupling. Thus the enlarged
symmetry forces all the six coupling constants appearing in (\ref{Gatto})
to become equal
\begin{equation} \label{ccs}
  g = g_H = g_K = g_H' = g_K' =  \lambda_V =\lambda_A
\end{equation}

The relations (\ref{ccs}) remain true, if the vector limit is
only broken by the gauge coupling of the light vector mesons.
However, it is known that there are other sources of vector symmetry
breaking, which are large and will affect the relation (\ref{ccs}).
Thus (\ref{ccs}) is likely to receive large corrections, but
it still may be useful for a first guess.

\section{Phenomenology}
In this section we shall discuss the phenomenological implications of
the combined heavy quark and vector symmetry. First we consider strong
decays. Off the vector limit the coupling constant $g$ is determined
from the strong decays $D^* \to D \pi$. The total rates for these decays
are given by
\begin{eqnarray}
\Gamma (D^{*+} \to D^0 \pi^+)& =& \frac{g^2}{96 m_{D^*}^3 \pi f^2}
\left[ (m_{D^*}^2 - (m_D + m_\pi)^2)
      (m_{D^*}^2 - (m_D - m_\pi)^2) \right]^{3/2} \\
\Gamma (D^{*+} \to D^+ \pi^0)& =& \frac{g^2}{192 m_{D^*}^3 \pi f^2}
\left[ (m_{D^*}^2 - (m_D + m_\pi)^2)
      (m_{D^*}^2 - (m_D - m_\pi)^2) \right]^{3/2} .
\end{eqnarray}
Experimentally only an upper bound for these decays is known
\cite{ACCMOR}
\begin{equation}
\Gamma (D^{*+} \to D^0 \pi^+) < 72 \mbox{ keV} (90\% \ CL)
\end{equation}
from which one obtains $ |g| < 0.63 $ \cite{chiral}.
In the vector limit one predicts that this coupling is the same for the
$D^* D \rho$ vertex, at least for the longitudinal components.
However, this vertex cannot mediate a real
decay due to phase space.

{}From the above Lagrangian one also predicts that the strong
decay $D_1(2420) \to D^{(*)} \pi$ is governed by the same coupling
constant $g$. In terms of this coupling constant we have
\begin{eqnarray}
\Gamma (D_1(2420)^0 \to D^+  \pi^-) &=& \Gamma (D_1(2420)^0 \to D^{*+} \pi^-)
\\ \nonumber
&=& \frac{g^2 m_{D^*}}{32 m_{D_1}^4 \pi f^2}
\left[ (m_{D_1}^2 - (m_D^* + m_\pi)^2)
      (m_{D_1}^2 - (m_D^* - m_\pi)^2) \right]^{3/2}
\end{eqnarray}
This decays has been seen, but no branching fraction has been
measured. Based on present data we have \cite{CLEOD}
\begin{equation}
\Gamma (D_1(2420) \to D^{(*)} \pi) < \Gamma_{tot} (D_1(2420))  =
20^{+7}_{-6} \mbox{ MeV}
\end{equation}
If we use the measured values for the masses, we extract a limit on $|g|$
which is stronger than the one from $D^{*+} \to D^0 \pi^+$, namely
$|g| < 0.35$. This, however, includes already symmetry breaking effects
due to the mass difference between the $D^*$ and the $D_1$. The result
depends on the third power of the pion three momentum $|\vec{p}_{cms}|$
in the cms frame, which is a factor of nine larger in the $D_1$ decay
than in the $D^*$ decay. The standard procedure to correct for phase
space effects is to divide out a factor $2|\vec{p}_{cms}|/m_D$, which
gives in the present comparison $|g| < 1.0$. However, this
is somewhat arbitrary; a comparison between the two results for
the decay constants needs to be more sophisticated and has to take into
account symmetry breaking.

The enlarged symmetry has also consequences for weak decays. For
the decay constants defined as
\begin{eqnarray}
  \langle 0 | \bar{q} \gamma_\mu (1-\gamma_5) Q_v | H_v (0^-) \rangle
   &=& i F_H m_H v_\mu \\
  \langle 0 | \bar{q} \gamma_\mu (1-\gamma_5) Q_v | K_v (0^+) \rangle
   &=& - i F_K m_K v_\mu
\end{eqnarray}
one obtains the prediction $F_H = F_K$ in the vector limit. This
prediction is somewhat counterintuitive, if one has a wave function
model for the mesons in mind. The $0^+$ and $1^+$ states are both
$P$-wave states in a wave function picture of the heavy-light meson,
while the $0^-$ and $1^-$ states are $S$-waves. Furthermore, in
such a picture the decay constants are proportinal to the wave
function at the origin, in other words, it should vanish for the
$P$ meson states. However, in real life this means that $F_H$ is
larger than $F_K$ and a difference between the two constants from
the point of vector symmetry has to be attributed to symmetry
breaking, which is known to be sizable.

Let us now consider semileptonic decays of heavy mesons into light
ones. In the vector limit, the transverse components of the light
vector particles decouple, and hence the decays into transversely
polarized light vector mesons should be suppressed compared to
the decay rate into longitudinally polarized light vectors.
Furthermore, the rate for longitudinally polarized vector mesons
should become equal to the rate into pseudoscalar mesons \cite{VLlong}.

However, this cannot be true for over the whole phase space
available, since the total rates
for semileptonic $D$ decays do not support this picture.
In table  \ref{tab1} we show the data for the ratio $\Gamma_L
/\Gamma_T$ for the decay $D \to K^* e \nu$ from  different
experiments. In the scenario considered above we would expect
$\Gamma_L / \Gamma_T \gg 1$, which is not consistent with the
measurements. In addition, the ratio of decay rates
\begin{equation}
R = \frac{\Gamma(D \to K^* e \nu)}{\Gamma(D \to K e \nu)} =
  =  \frac{\Gamma_T + \Gamma_L} {\Gamma (D \to K e \nu)}
\end{equation}
is experimentally $R = 0.51 \pm 0.18 $ in the neutral $D$ decays
and $R = 0.74 \pm 0.19$ in the charegd $D$ decays. From the vector limit
one would conclude that $R \sim 1$, with the real value being less than
unity due to the small ratio $ \Gamma_L / \Gamma (D \to K e \nu) $, but
even for the measured values
of $\Gamma_L / \Gamma_T \sim 1.2$ the ratio $R$ is inconsistent
with the above picture.

\begin{table}
  \begin{center}
  \begin{tabular}{|c||c|c|c|}
     \hline
     & E653 \cite{E653} & E691 \cite{E691} & MARK III \cite{MARK} \\
     \hline  \hline &&& \\
    $\Gamma_L/\Gamma_T$ & $1.18 \pm 0.18 \pm 0.08$ &
    $1.8^{+0.6}_{-0.4} \pm 0.3$ & $0.5 ^{+1.0 +0.1}_{-0.1-0.2}$ \\&&& \\
    \hline
  \end{tabular}
  \caption{Measurements of $\Gamma_L/\Gamma_T$ in $D \to K^* \ell \nu$.}
  \label{tab1}
  \end{center}
\end{table}

However, chiral symmetry is expected to be valid only for
sufficiently soft light particles. This assumption is not valid
over the whole phase space in semileptonic heavy to light decays.
Neglecting the mass of the light particle, the maximal energy of the
light meson is $E_{max} = m_H / 2$, which becomes large in the
heavy mass limit. The description based on the chiral limit is
certainly valid only close to the kinematic point, where the
energy of the light particle is small, $E \sim \Lambda_{QCD}$,
and the problems dicsussed above for the total rate may be related
to the inadequacy of the chiral limit in most of the phase space.
In order to test this one has to compare the lepton spectra of
the decays into pseudoscalar and vector mesons.

Defining the form factors for the semileptonic decays according to
\begin{eqnarray}
\langle K(p) | \bar{s} \gamma_\mu (1-\gamma_5) h_v | H(v) \rangle
&=&
F_+ (m_H v_\mu + p_\mu)
\label{Fplus} \\
\langle K (p,\varepsilon) |
	\bar{s} \gamma_\mu (1-\gamma_5) h_v | H(v)\rangle
&=&
m_H \varepsilon_\mu  F_1^A
+ m_H (v \cdot\varepsilon) v_\mu  F_2^A
+ i \epsilon_{\mu\nu\rho\sigma}
	\varepsilon^\nu v^\rho p^\sigma  F^V
\label{F1A}
\end{eqnarray}
where form factors proportional to $q_\mu = m_H v_\mu - p_\mu$
have been omitted.
Neglecting the mass of the light meson one obtains for the
differential rates
\begin{eqnarray}
\frac{d\Gamma_0(D^0 \to K^+ e \nu)}{dz}
&=&
    \frac{G_F^2 m_D^5}{192 \pi^3} |V_{cs}|^2
     z^3 |F_+|^2
\\
\frac{d\Gamma_L(D^0 \to K^{+*} e \nu)}{dz} &=&
  \frac{G_F^2 m_D^5}{192 \pi^3} |V_{cs}|^2
  \frac{1}{4 r_*^2} z^3
  \left[F_1^A +
       \frac{z}{2} F_2^A \right]^2
\\
\frac{d\Gamma_T(D^0 \to K^{+*} e \nu)}{dz} & = &
  \frac{G_F^2 m_D^5}{192 \pi^3} |V_{cs}|^2
  (1 - z ) z
  \left[ 2 (F_1^A)^2 +
       \frac{z^2}{2}  (F^V)^2
  \right]
\end{eqnarray}
where $z = 2 (vp)/m_H$. Note that
in the longitudinal rate the dependence on $r_* = m_H / m_{K^*}$
has to be kept, since  the longitudinal rate behaves as $1/m_{K^*}^2$.
Approaching the vector limit, the longitudinal rate has to have a
finite limit and one obtains the following relation between the form
factors \cite{VLlong}
\begin{equation}
m_D F_1^A + (v \cdot p) F_2^A = 2 g_V f F_+ .
\end{equation}
Together with the form factor $F_+$ which may be derived from the
chiral Lagrangian \cite{chiral}
\begin{equation}
F_+ = - \frac{F_D}{2F_K}
      \left( 1 + g \frac{m_D - v \cdot p}{v \cdot p + m_{D_s} - m_D}
      \right)
\end{equation}
one obtains a prediction for the shape of the spectra,
which should hold for not too large
energies of the light meson.

Finally, it is worth to mention that the implications of vector
symmetry for nonleptonic weak decays of heavy mesons have also been
considered \cite{VLHQ,VLHQ1,VLHQ2}, with the result that the relations
between the various decay rates are within a factor 2 in agreement
with data.

\section{Conclusions}
Combining heavy quark and chiral symmetry in its conventinal form
has lead to many interesting predictions. However, in the
conventional formulation of the chiral Lagrangian for the light
degrees of freedom only the pions appear, but for phenomenological
applications it is desirable to have also the light vector mesons
in the Lagrangian for the light degrees of freedom.

{  The light vector mesons may be introduced by writing down all coupling
terms allowed by chiral and Lorentz invariance; this, however, leads to
a proliferation of unknown coupling constants in the heavy-meson chiral
Lagrangian, which need to be fixed from experimental input.

If chiral symmetry is indeed realized in an unbroken way and the vector
limit may be used as a starting point, the symmetry becomes larger than
in the conventional picture. }
Although it is still obscure, whether and how this enlarged
symmetry is generated from QCD, it is still a useful tool to
reduce the number of independent coupling constants once the
light vector mesons are introduced.

Vector symmetry leads to a relation between the matrix elements
involving light pseudoscalars and longitudinal components of light
vector mesons. However, for the heavy particles, vector symmetry
has to be realized in a different way, since the vector limit
corresponds to the massless limit for the light mesons.
We have chosen to have an explicit representation of the unbroken
symmetry for the heavy mesons, such that we have degenerate
parity doublets for the heavy states. We have included the heavy
ground state spin symmetry doublet for the mesons and a spin
symmetry doublet of excited, positive parity mesons, which we
identified with the chiral partners of the ground state spin
symmetry doublet. Due to the enlarged symmetry, the Lagrangian
with this set of fields still has only one unknown coupling
constant, which is the same as the one appearing in the ususal
chiral Lagrangian and which is related to the $H^* H \pi$
vertex.

The comparison of this idea with phenomenology shows that
large symmetry breaking will be present. The symetry breaking effects
have been parameterized for the light degrees of freedom \cite{Cho,VLtau}
and a similar
approach may be taken for heavy light systems as well.
However, a more detailed
comparison in order to determine the size of symmetry breaking
has to wait until better data become available.

\end{document}